**Accepted version, published in the *Journal of International Financial Markets, Institutions and Money*, 2017, vol. 50, pp. 219-234.**
**DOI:** https://doi.org/10.1016/j.intfin.2017.08.012.

# PRICING AND USAGE:
# AN EMPIRICAL ANALYSIS OF LINES OF CREDIT

Miguel A. Duran[*]

**Abstract:** The hypothesis that committed revolving credit lines with fixed spreads can provide firms with interest rate insurance is a standard feature of models on these credit facilities' interest rate structure. Nevertheless, this hypothesis has not been tested. Its empirical examination is the main contribution of this paper. To perform this analysis, and given the unavailability of data, we hand-collect data on usage at the credit line level itself. The resulting dataset enables us also to take into account characteristics of credit lines that have been ignored by previous research. One of them is that credit lines can have simultaneously fixed and performance-based spreads.

## 1. Introduction

Committed revolving credit facilities are a major source of corporate financing (Sufi 2009). A large fraction of them are provided by syndicates of lenders in which the participation of banks and other finance institutions from different countries is common (Carey et al. 1998, Carey and Nini 2007). Loan syndication is indeed an important means to channel funds internationally (Xu and La 2015).

According to models on the interest rate structure of revolving lines of credit, one of the reasons for the demand of these loan commitments is that they can provide firms with interest rate insurance (Campbell 1978, Thakor et al. 1981). From the borrowers' perspective, the protection provided by credit lines can be twofold: against the volatility of general market interest rates, such as the London Interbank Offered Rate (LIBOR), and against increases in firms' borrowing costs due to poor performance. Fixed rate agreements hedge both types of risk, whereas fixed spread commitments (i.e., those where the interest rate on drawdowns is the sum of a benchmark rate plus a fixed spread) protect against rises in firms' spot risk premiums. However, no interest rate protection is expected to be provided under performance-sensitive credit lines, since they are designed to make the interest rate on revolving loans close to the spot rate.

---

[*] I thank Anthony Saunders for helpful comments. This work was supported by the Spanish Ministry of Economics and Competitiveness (Project ECO2014-52345-P).

The view of revolving credit lines as interest rate hedging instruments suggests that firms whose credit quality deteriorates behave differently depending on the type of interest rate structure. A higher credit risk increases the firm's spot risk premium. Therefore, if the firm has a fixed rate or spread commitment, drawing down from the credit line may eventually become relatively cheaper and, hence, the firm would use the facility more intensively. In contrast, if the interest on revolving loans is equal to a benchmark rate plus a performance-sensitive markup, usage is not expected to increase when the firm is financially distressed, since there would be no significant difference between the spot and commitment prices.

The empirical analysis of these differences across interest rate structures is the main aim of this paper; that is, leaving fixed rate credit lines aside, which are infrequent, our central question is whether credit line usage is affected by their interest rate structure as corporate creditworthiness deteriorates. To the best of our knowledge, this question has not been previously investigated. Examination of the relation between corporate creditworthiness and usage is the closest previous research has come to our approach. Indeed, Berg et al. (2016) find a negative relation between firms' credit risk and facility usage, but their analysis is conducted at the firm level, with no reference to the interest rate structure. However, the attempt to test whether credit lines provide interest rate insurance actually requires both performing the analysis at the facility level and taking into consideration pricing schedules. If data on credit line usage at the line level are not employed, the interest rate structure of a concrete credit line cannot be linked to how this facility is used. Indeed, the unavailability of data on usage at the credit line level is probably one of the reasons why the relation between interest rate structures and usage has not been previously studied. The reason why interest rate structures must be taken into account is also straightforward: Not all interest rate structures provide interest rate protection.

To perform our analysis, nevertheless, we must refine the taxonomy of interest rate structures. In particular, the standard, dichotomous separation between credit lines with either fixed or variable spreads must be qualified. In this sense, no previous academic work has paid attention to a common feature of credit lines: Most (about 90% in our sample) offer the borrower the choice between two different interest rates that differ in whether the benchmark rate is LIBOR or, as it is commonly called in credit contracts, an alternate base rate (ABR). As Wight et al. (2007: 224) put it, "A borrower can elect to borrow either [alternate] base rate loans or LIBOR loans, or both simultaneously, under a credit agreement, and the borrower can switch back and forth between the two types of loans." According to our data, the most common categories of interest rate structures are xp (the spreads on ABR and LIBOR loans are fixed and performance sensitive, respectively), xx (the spreads on ABR and LIBOR loans are both fixed), and pp (the spreads on ABR and LIBOR loans depend on corporate performance).

Based on this classification, we can now be more specific regarding our main research aim. According to the view of fixed spread credit lines as providers of interest rate insurance, we examine whether creditworthiness-deteriorating firms use their revolving lines more intensively if the latter have xp or xx interest rate structures. However, since performance-sensitive pricing does not hedge against rising risk premiums, we do not expect to observe a similar effect for pp lines.

Our analysis focuses on a random sample of US publicly traded corporations included in the Standard & Poor's (S&P) Compustat database that, according to Thomson Reuters Loan Pricing Corporation's DealScan database, have at least one active credit line in the sample period, which goes from 2006:Q1 to 2012:Q2. Nevertheless, the requirement to use data on credit line usage at the level of the facilities themselves raises two main problems. On the one hand, the only commercial database that provides information on credit line usage is the S&P Capital IQ database, but it does so at the firm level. Therefore, we use 10-Q and 10-K US Security and Exchange Commission (SEC) filings to build a unique dataset on the quarterly usage of credit lines at the facility level. On the other hand, large firms that use credit lines tend to have a considerable number of them simultaneously and they usually disclose information about usage in an aggregate manner. To overcome this problem, we focus on firms with assets below $20 billion, which rarely have more than two revolving credit facilities active at the same time and usually report them separately. The $20 billion threshold implies that our sample firms are comparable, in terms of asset size, to mid- and small-cap firms.

Besides collecting data on credit line usage from SEC filings, we use the agreement contracts uploaded in the Electronic Data Gathering, Analysis, and Retrieval (EDGAR) system to obtain information on credit facilities that was incomplete or missing in DealScan. The search process to collect these contracts reveals a relevant limitation of DealScan: This database does not include a relatively large number of amendments to credit agreements.

Our results show that the relation between corporate financial distress and facility utilization is positively affected by xp lines. This finding suggests that this type of facility protects firms against increases in corporate spot risk premiums generated by deteriorating quality; that is, it supports the hypothesis that fixed spread credit lines provide firms with interest rate insurance. Accordingly, risk-increasing borrowers with xp lines appear to use them to shift risk to their lenders (Jensen and Meckling 1976, Duran and Lozano-Vivas 2014a, 2014b). Such risk shifting could help explain a striking finding: At the start of the 2008 crisis, the relative frequency of xp lines began a continuous decline that has led to a situation where this type of interest rate structure seems to have virtually vanished from the market.

Regarding xx lines, our empirical evidence indicates that this pricing schedule causes no differential effect on how facility usage relates with corporate creditworthiness, that is, the way in which xx lines are used seems to contradict the hypothesis that fixed spread credit lines provide a hedge against fluctuations in corporate risk premiums. We also find that having an xx line reduces usage. The differences between our results for xp and xx highlight that the standard dichotomy between fixed and variable spreads is insufficient to analyze the interest rate structure of credit lines.

In tune with our probit analysis of the variables that affect the probability of observing the different interest rate structures, our results for xx lines seem to be caused by lenders resorting to xx lines to protect themselves from borrowers with deteriorating quality. In this sense, having no S&P corporate credit rating or being rated speculative grade substantially increases the probability of having an xx line. Our probit estimation also helps us to address potential endogeneity concerns.

The remainder of the paper proceeds as follows. Section 2 reviews the literature and puts forward the hypothesis to be tested. Section 3 describes the data and summary statistics. Section 4 discusses the empirical results. Section 5 concludes the paper. The Appendix defines the variables used in the analysis.

**2. Previous literature, theoretical motivation, and hypothesis**

The theoretical motivation for this paper has its roots in the literature that views revolving loan commitments as a risk-sharing mechanism where lenders protect borrowers against potential instability in financing costs. An early work in this field is that by Campbell (1978), who shows that fixed spread commitments provide the borrower with full insurance against rises in the latter's credit risk premium. In addition, consistently with what is observed in the market for credit lines, Campbell's (1978) model predicts that different types of interest rate structures are expected to coexist in the market. In a similar vein, Thakor et al. (1981) identify fixed spread loan commitments as put options: By entering into one of these commitments, the borrower purchases the right to sell risky indebtedness to the lender at a specified price and, thus, the borrower hedges against the variability of the spot rate. Therefore, a firm will increase the usage of fixed spread credit lines as its quality deteriorates and, consequently, its spot risk premium will rise. This option-based view of fixed spread credit lines also underlies the analysis of Hawkins (1982) and Boot et al. (1987).

According to this theoretical approach, we should expect the levels of protection and, hence, usage behavior to not be homogeneous across types of spreads. Indeed, not all pricing schedules embed option-like protection against increases in credit risk premiums: Performance-sensitive spreads are expected to adjust to these increases and, therefore, prevent borrowers from receiving what Shockley and Thakor (1997) describe as a risk premium subsidy. Therefore, drawdowns on lines with performance-based pricing are not expected to increase as firms' financial distress rises. In contrast, for fixed spread credit lines, we should expect a relatively more intense usage if firms become more financially distressed.

Nevertheless, as Wight et al. (2007) point out, despite no previous academic work having taken it into consideration, a major fraction of credit lines contains pricing schedules that allow firms to borrow at different rates, with xp, xx and pp being the most widespread interest rate structures. Accordingly and given that xp and xx include fixed spreads, we test the following hypothesis.

HYPOTHESIS: Lines of credit with xp or xx pricing schedules provide holding firms with interest rate insurance, whereas pp facilities do not.

To test this hypothesis, we expect the effect of financially distressed firms on facility usage to be more positive for firms with xp or xx lines but not for pp facilities.

From an empirical perspective, previous works analyze the relation of facility pricing with different characteristics of lenders, borrowers, and the facilities themselves, but no previous research examines whether interest rate structures insure borrowers as they lose creditworthiness. Moreover, probably due to the unavailability of data, these previous works mainly focus on the determinants or effects of contractual pricing provisions at a concrete moment, usually at origination.

A pioneering contribution to the empirical examination of credit line pricing is that of Shockley and Thakor (1997). For a 1989–1990 sample of facilities purchased by publicly traded firms, they find empirical support for the hypothesis that multiple fee structures should not be observed for well-known, high-quality firms. Strahan (1999) shows that the credit agreements of borrowers that are riskier at origination include higher initial all-in spreads. An early paper on performance-sensitive pricing is that by Asquith et al. (2005), who study the potential cost-saving effects of this type of pricing in both term and revolving credit agreements. Following this line of research, Ivanov (2012) points out that the primary role of performance pricing in bank debt contracts is to delay costly renegotiation. Manso et al. (2010) show that borrowers with performance-sensitive loans are more likely to be upgraded one year after origination than borrowers with fixed-interest loans are. Similar results are obtained by Begley (2012). Focusing on the effects of the 2008 crisis, Santos (2011) examines whether banks with larger losses increase the spreads that they charge at their credit lines' inception. Tchistyi et al. (2011) find empirical evidence supporting that performance-sensitive debt can be the source of a potential conflict of interest between firms' stockholders and managers. Among other hypotheses, Berg et al. (2016) examine whether there is a negative relation between facility usage and a firm's credit quality. Nevertheless, the hypothesis that fixed spread facilities provide interest rate insurance is not among those that these authors test; indeed, they do not take into account interest rate structures and perform the analysis at the firm level.

Examining the view of revolving facilities as a hedging instrument is this paper's main contribution to the literature on the market for credit lines. In addition, to the best of our knowledge, no previous academic work has analyzed the complex nature of credit lines' interest rate structures. Thus, our research can also contribute to shed light on this unexplored feature of revolving facilities.

## 3. Data and sample characteristics

This section describes the sample construction process, as well as the basic characteristics of the data. Given the novelty of our approach to credit lines' interest rate structures, we discuss their main features in the last subsection.

**3.1. The DealScan–Compustat dataset**

To build our final dataset we merge information from three sources: DealScan, quarterly accounting data from Compustat, and a manually collected dataset. Using DealScan and Compustat enables us to obtain a dataset with the quarterly accounting data of firms that have at least one outstanding revolving credit line in the sample period. The hand-gathered dataset provides information about credit agreements that is not or is insufficiently covered by DealScan.

Our analysis requires firms' accounting information and, therefore, we exclude firms in DealScan that cannot be matched to those in Compustat. In addition, we only keep observations corresponding to US non-financial firms and to dollar-denominated revolving credit loans active in the sample period,[1] from 2006:Q1 to 2012:Q2, that is, through 26 quarters.[2] The end of the sample period is defined by the dataset of Chava and Roberts (2008), which enables the merging of the DealScan and Compustat databases. The analysis is quarterly so that the results better reflect the high frequency of credit agreement renegotiations (Roberts and Sufi 2009, Roberts 2015).

A considerable number of large corporations have more than one credit line active simultaneously and usually report on facility usage in an aggregate manner. Since we require data on usage at the credit line level, this aggregate information is inadequate for our research goal. Accordingly, to make it possible to hand-collect disaggregated data on drawdowns, we exclude from our DealScan–Compustat dataset firms with an asset book value above $20 billion in any sample quarter. The reason for this concrete threshold is that the maximum asset value of a firm included in the S&P MidCap 400 or SmallCap 600 during the sample period amounts to $19.921 billion, which was reached by Republic Services Inc. in 2008:Q4. Therefore, our sample consists of companies that are not necessarily part of these stock indexes and, therefore, any bias associated with being listed is avoided; however, in terms of maximum asset size, the sample companies resemble those listed in these indexes. In relation to how the $20 billion threshold affects the Compustat universe, the large majority (96.37%) of non-financial US companies included in this database in 2006:Q1–2012:Q2 have an asset size below this threshold during the whole period.

---

[1] Following Berg et al. (2015), we select loan commitments whose DealScan variable *loantype* is either *Revolver/Line < 1 Yr.*, *Revolver/Line >= 1 Yr.*, *364-Day Facility*, *Limited Line*, or *Revolver/Term Loan*.
[2] However, our analysis of the probabilities of observing the different interest rate structures requires Compustat data from 2001:Q2 onward.

Following Sufi (2009), we also require firms to have a minimum number of consecutive quarters in Compustat and with active lines of credit.[3] The resulting dataset has 206,883 facility–quarter observations, with 8,908 lines of credit and 2,545 firms. The next step is to randomly select 150 firms to serve as the basis for the manual process of data collection.

**3.2. The manually collected dataset**

According to SEC regulations, public firms must provide detailed information about their credit lines in 10-Q and 10-K reports (Kaplan and Zingales 1997, Sufi 2009). These reports reflect a significant limitation of DealScan: Firms refer to amendments to existing credit lines and occasionally to newly originated credit lines that DealScan does not include. Therefore, the information provided by this database could be insufficiently accurate for analyzing how credit line characteristics affect variables that change over time, such as credit line usage. To overcome this limitation, we search for information about new lines or amendments not covered by DealScan in the body of 10-Q and 10-K reports and in the list of exhibits that appears at the end of these reports. Reference to an exhibit in this list is usually complemented by information that allows us to locate the original contract in EDGAR. Following Roberts and Sufi (2009), once we find a contract, we include it in our dataset if it refers to a new credit line or does not leave unchanged the principal, interest rates on drawdowns, fees, or maturity.

The proportion of credit lines that we have manually added is almost a third (30.72%) of the facilities in our dataset. The vast majority of these lines were amendments and amended and restated agreements (89.45%). These percentages raise concern over how DealScan follows the renegotiations of credit lines that it covers at origination.

---

[3] To describe this condition, let us define the concepts *starting quarter* and *ending quarter* by means of an example. Consider a firm that is included in Compustat between 2006:Q2 and 2010:Q2 and that has two lines of credit active in the sample period 2006:Q1 to 2012:Q2. The origination (maturity) quarters of these lines are 2006:Q3 (2007:Q3) and 2007:Q2 (2103:Q2), respectively. In this case, the starting quarter is the latest between 2006:Q2 or 2006:Q3, that is, the latest quarter between the first quarter in which the firm is included in Compustat (2006:Q2) and the earliest quarter among the origination quarters of the facilities the firm has in the sample period (2006:Q3 between 2006:Q3 and 2007:Q2). If the latest quarter between the candidates 2006:Q2 and 2006:Q3 is earlier than 2006:Q1, the starting quarter will be 2006:Q1. The ending quarter is the earliest between 2010:Q2 and 2013:Q2, that is, the earliest quarter between the last quarter in which the firm is included in Compustat (2010:Q2) and the latest quarter between the maturity quarters of the facilities the firm has in the sample period (2013:Q2 between 2007:Q3 and 2013:Q2). If the earliest quarter between the candidates 2010:Q2 and 2013:Q2 is later than 2012:Q2, the end quarter will be 2012:Q2. Given these definitions, we require firms to have at least four consecutive quarters between the starting and ending quarters. This condition is less restrictive than Sufi's (2009), which requires firms to have at least four consecutive years (in an eight-year period) of positive data on the main variables of analysis. It is established to reduce the probability of credit lines with no observations in the final randomly chosen dataset.

In addition, we collect the contracts of sample facilities covered by DealScan. Thus, we obtain the agreements of all the credit lines included in our dataset (except for four, which were not found) and we can gather information about features of credit lines not or insufficiently covered by DealScan.

Forms 10-K and 10-Q also provide information on credit line usage at the credit line level. Nevertheless, despite regulations that imply that firms must report their available credit lines, there is no explicit requirement for disclosing credit line usage. Therefore, we drop any quarter–facility observation for which we have not found data on usage. Similarly, we drop facility–quarter observations for which, despite the $20 billion threshold on firm asset size, a company reports usage of its credit lines in an aggregate manner.

As Roberts and Sufi (2009) point out, DealScan's coverage of data on interest rate structures has limitations. Accordingly, we use credit contracts to check the information that DealScan provides on the types of base rates and spreads, the values of the spreads if they are fixed, and the maximum and minimum values of the spreads if they are variable. Data on manually added credit lines are directly obtained from credit contracts. Once data on interest rate structures are ready to be used, we restrict the sample to facility-–quarter observations for which available information allows us to determine the type of interest rate structure, that is, the types of base rates that borrowers can select and the nature of the spreads. This condition yields our final sample: an unbalanced panel of 2,595 facility–quarter observations that includes 586 lines of credit, which is the basic unit of observation in our analysis. Although 150 firms were randomly selected from the DealScan–Compustat, this number is reduced to 139 in the final dataset.

In relation to the rest of the credit line characteristics, we proceed in a similar manner to what we have done for interest rates; that is, for manually added credit lines, we hand-collect data on fees, principal, maturity, method of syndication, and whether these lines are secured. For the sample credit lines included in DealScan, we check the data that it provides on these characteristics with credit contracts. Besides reducing the probability of potential errors, this comparison avoids situations where, for instance, credit line usage is larger than the principal of the facility.

Regarding fees, we focus on the four most common: commitment, annual, upfront, and utilization fees. Data on fees allow us to control for other costs associated with credit lines. Although we take into account upfront fees, on the one hand, they are generally provided for in nonpublic fee letters instead of in credit agreements (Wight et al. 2007).

On the other hand, Berg et al. (2016) suggest that DealScan provides more information related to upfront fees than is actually available in credit agreements. Therefore, we proceed as follows. Although a credit contract makes no reference to an upfront fee, we consider that the corresponding credit line includes such a fee if DealScan does. In addition, we test the robustness of our results if upfront fees are not taken into account as regressors.

DealScan includes the variables *primarypurpose* and *secondarypurpose*. These variables seem to refer to what would be a facility's main and auxiliary purposes, respectively. However, typical credit agreements do not establish any priority among purposes and frequently include more than two.[4] Accordingly, we hand-collect data on credit line purposes. Only 18.43% of the sample credit lines have a single purpose and 34.12% have two. To use this information in our analysis, we follow Ivashina and Scharfstein (2010a): We split credit lines between those that could be used for corporate restructuring (leveraged buyouts, mergers and acquisitions, and stock repurchases) and the remainder.

We also manually collect data on two additional features of credit lines that are not covered by DealScan; specifically, we gather data on violations of financial covenants and whether credit lines can be used to support the issuance of letters of credit.

### 3.3. Sample general statistics

Manually gathered information about variables available in DealScan is used to check for differences between this database and credit agreements. Very few credit contracts differ from DealScan's data on line principal (21 lines, 5.95% of the sample lines included in DealScan), maturity (19 lines, 5.38%), and whether lines are syndicated (one line, 0.28%) or secured (six lines, 1.70%).

Regarding the characteristics of sample facilities, Table 1 displays summary statistics other than interest rates. These statistics are calculated as if there were one observation per credit line, with two exceptions: The figures for the usage-to-principal ratio are based on the whole sample and those of covenant violations are calculated over firms. Approximately a quarter (24%) of credit line commitments are, on average, drawn down.

---

[4] For instance, the credit agreement of Pacific Sunwear of California Inc., dated September 14, 2005, states, "The proceeds of the Loans will be used only to refinance certain existing credit facilities of the Borrower and to finance the working capital needs, capital expenditures, acquisitions (including Permitted Acquisitions), dividends, distributions and stock repurchases, and for general corporate purposes of, the Borrower and its Subsidiaries." For more detailed information, see Duran (2017).

A large majority of the sample credit lines (86%) support the issuance of letters of credit. Corporate restructuring is among the purposes of about half (45%) of the sample credit lines. The average sample facility has a stated maturity slightly shorter than 42 months and a principal of approximately $250 million. In addition, 92% and 63% of the sample lines are syndicated and secured, respectively. The rate of firms violating covenants in our dataset is 21%, lower than the percentages of Chava and Roberts (2008) and Sufi (2009). A possible anecdotal evidence-based explanation for this result is that firms have learned to avoid financial covenant violations through renegotiation as the use of credit lines has become increasingly widespread. In this sense, 10-Q and 10-K filings suggest that firms that anticipate a covenant breach often request their lenders to renegotiate credit conditions.[5] Indeed, although Roberts (2015) does not differentiate between renegotiations due to a covenant violation or in anticipation of one, the author points out that these are the motives of approximately 28% of renegotiations in his sample.

**[Table 1]**

Table 1 also presents statistics for revolving credit lines included in DealScan, denominated in US dollars, active in the sample period, and made to non-financial US firms. These lines are quite similar to the sample ones, although the latter have a slightly lower average maturity and principal (eight months shorter and $16 million less, respectively). The percentage of syndicated and secured lines is also lower in the sample than in DealScan (7% and 16% less, respectively).

**[Table 2]**

Table 2 shows summary statistics for sample firms and non-financial US firms in Compustat during the sample period. To reduce the effect of outliers and coding errors, we winsorize all financial variables from Compustat at the top and bottom percentiles in both our sample and the Compustat population. Relative to the latter, the former contains firms that are, on average, smaller, have lower market-to-book and liquidity ratios, and are less highly levered. The sample firms also have relatively more tangible assets, and are more profitable. Altman's (1968) Z-score suggests that the firms in our dataset are less financially distressed. In addition, rated firms and firms with a speculative-grade rating are more common in our sample than in Compustat. Overall, these differences are largely consistent with the sample selection criteria that we use, particularly under the conditions that require firms to have an asset size below $20 billion and to hold at least one outstanding credit line in the sample period.

---

[5] The reports of COMPX International Inc. are an example of renegotiation in anticipation of a covenant violation. In the 10-Q report of 2009:Q2, this firm announced, "It is probable that we will not be able to comply with the interest coverage ratio covenant as of September 30, 2009.… We have begun discussions with the lenders to amend the terms of the existing credit facility to, among other things, modify the interest coverage ratio covenant and change the amount of the consolidated net worth covenant." In September 2009, the firm filed an amendment and in the 10-Q report of 2009:Q3 it stated, "The primary purpose of the Third Amendment was to adjust certain covenants … in the Credit Agreement in order to take into consideration our current and expected future financial performance."

### 3.4. Interest rate structures

As Wight et al. (2007) point out, a large percentage of credit agreements offer borrowers the choice between two types of loans and borrowers usually even have the chance to convert one type to the other. The type is mainly defined by the base rate used to compute interest rates and the base rate can be LIBOR or ABR. In our sample, 90.77% of the lines offer borrowers the choice between LIBOR or ABR loans.

Margins added to the base rates may be variable or fixed. We classify variable spreads in three categories: performance sensitive, usage based, and maturity based. In our sample, the performance-sensitive category corresponds to the criteria that DealScan identifies as the ratio of total debt to cash flow, the senior debt rating, the fixed-charge coverage ratio, the ratio of senior debt to cash flow, the ratio of debt to tangible net worth or the leverage ratio, the ratio of senior debt to cash flow, and the senior leverage ratio. Sample facilities also include two performance-sensitive criteria that cannot be matched to any DealScan category; specifically, the spreads on ABR and LIBOR loans are tied to the borrowers' credit default swap spreads in four lines and depend on the equity-to-assets ratio in one line. Among the DealScan criteria covered by our performance-sensitive category, the most common is the ratio of total debt to cash flow, followed by the senior debt rating. Our usage-based category comprises three of DealScan's criteria: availability, the borrowing base, and balance outstanding. The last of our categories is the criterion that DealScan identifies as maturity and refers to spreads that depend on the time remaining to the termination date.

[Table 3]

Panel A of Table 3 shows the distribution of sample lines across bases rates and spreads. Performance-sensitive spreads are the most common for both ABR and LIBOR loans; specifically, 49.82% and 70.38% of lines that allow ABR and LIBOR borrowings, respectively, have performance-sensitive spreads. This type and fixed spreads account for the great majority of sample lines: Almost 90% of the spreads on either ABR or LIBOR loans depend on corporate performance or are fixed. The rest of the lines have spreads determined by credit line usage, maturity, or multiple criteria. As Panel B indicates, there are three possible situations in which the samples lines have spreads determined by more than one criterion: Spreads depend on (a) credit line usage and corporate performance (usg & prf in Panel B); (b) credit line usage and time to maturity (usg & mat), or (c) corporate performance, credit line usage, and time to maturity (usg & prf & mat). The total number of credit lines with multiple criteria for either LIBOR loans or both ABR and LIBOR loans is relatively low (2.05% over all sample lines).

Given that most credit lines include the choice of borrowing either ABR or LIBOR loans, we group lines of credit depending on the spreads on both types of base rates. We thus obtain five different interest rate structures, named according to the following simple convention. We use two letters: The first refers to spreads on ABR loans and the second to spreads on LIBOR loans. There are four letters, as many as types of spreads: $p$ for performance-sensitive spreads, $x$ for fixed spreads, $u$ for spreads determined by credit line usage, and $m$ for spreads that depend on time to maturity. In line with this nomenclature, the interest rate structures in our sample are xp, xx, pp, uu, mm, and xu. The interest rate structure xp, for instance, has a fixed spread on ABR borrowings and a performance-sensitive spread on LIBOR loans. If a credit line has only one type of base rate, it is included in the group where the spreads of ABR and LIBOR loans are equal to each other and match that credit line's type of spread. For example, a line with a fixed spread on LIBOR loans and no option to borrow ABR loans is included in the group xx.

As Panel B in Table 3 shows, the most common interest rate structure among sample lines is pp (48.81%), followed by xx (20.48%) and xp (19.62%). Hence, these three types cover almost 90% of sample credit lines. Slightly below 8% of the sample lines are of type uu, whereas the percentage of those of type mm or xu is insignificant.[6]

## 4. Effect of interest rate structure on credit line usage

In this section, we present the empirical analysis of the relation between interest rate structures, credit line usage and creditworthiness. Specifically, we test the hypothesis that fixed spread credit lines, that is, xp and xx facilities, hedge against increases in spot risk premiums.

---

[6] Two credit lines do not match any of the interest rate structures shown in Panel B of Table 3. One of them is the only fixed rate credit line found in our sample. The credit agreement of the other credit line does not make any reference to interest rates paid on drawdowns.

## 4.1. Empirical model

Our base empirical model is

$$Usage\ ratio_{i,t} = \alpha_0 + \alpha_1 \cdot xp_i + \alpha_2 \cdot xx_i + \alpha_3 \cdot pp_i + \alpha_4 \cdot zs_{i,j,t-1} + \alpha_5 \cdot xp_i \cdot zs_{i,j,t-1} + \alpha_6 \cdot xx_i \cdot zs_{i,j,t-1} + \alpha_7 \cdot pp_i \cdot zs_{i,j,t-1} + \sum_{m=1} \alpha_{m,8} \cdot H_{m,i} + \sum_{n=1} \alpha_{n,9} \cdot K_{n,i,j,t-1} + \alpha_{10} \cdot Crisis_t + \alpha_{11} \cdot Violation_{i,j,t} + \omega_j + \varphi_t + \gamma_h + \varepsilon_{i,t},$$

where, for quarter $t$ and facility $i$ entered into by firm $j$ and lender $h$, $Usage\ ratio_{i,t}$ is the rolling four-quarter average of the ratio of credit line usage to principal; $xp_i$, $xx_i$, and $pp_i$ are (0, 1) indicator variables for the three main interest rate structures;[7] $zs_{i,j,t-1}$ is the lagged Z-score; $H_{m,i}$ is the $m$th control variable of credit line characteristics; $K_{n,i,j,t-1}$ is the $n$th lagged control variable of corporate characteristics; $Crisis_t$ is a (0, 1) indicator variable for the 2008 crisis period; $Violation_{i,j,t}$ is a (0, 1) indicator variable for whether firm $j$ is in violation of financial covenants of credit line $i$; $\omega_j$, $\varphi_t$, and $\gamma_h$ are one-digit Standard Industrial Classification (SIC) code industry, quarter, and lead lender fixed effects, respectively;[8] and $\varepsilon_{i,t}$ is a random error term that is assumed to be correlated within credit line observations and potentially heteroskedastic. To take into account this assumption, we use ordinary least squares regressions clustered at the credit line level.

Following Sufi (2009), we use Altman's (1968) Z-score to capture the level of corporate financial distress. Accordingly, the parameters of the interaction terms between the Z-score and the indicator variables for the interest rate structures ($\alpha_5$, $\alpha_6$, and $\alpha_7$) are the key coefficients to test the hypothesis that fixed spreads provide interest rate insurance. Since the Z-score is an inverse measure of financial distress, empirical evidence supports our hypothesis if $\alpha_5$ and $\alpha_6$ are negative whereas $\alpha_7$ is not. In this regard, a negative and significant coefficient of the interaction term between the Z-score and, for instance, xp ($\alpha_5$) indicates that the effect of corporate financial distress on facility usage is more positive for xp credit lines.

The control variables that we use are standard firm and line characteristics likely to affect credit line usage (e.g., Sufi 2009, Berrospide et al. 2012, Berg et al. 2016, Delis et al. 2016). Our analysis takes into account eight corporate features as control variables: (1) size, measured as the natural logarithm of assets; (2) investment opportunities, measured as the market-to-book ratio; (3) asset tangibility, measured as the ratio of tangible assets to total assets; (4) profitability, measured as the ratio of earnings before interest, taxes, depreciation, and amortization to assets; (5) book leverage, measured as the ratio of total debt from the balance sheet to assets; (6) liquidity, measured as the ratio of current assets to current liabilities; (7) the risk of default, measured by whether a firm is rated investment grade or speculative grade; and (8) information asymmetry, measured by whether firms are not rated and could therefore face higher financing costs resulting from lack of transparency and tougher access to credit markets.

---

[7] The omitted category refers to the rest of credit lines, for example, those that have spreads determined by credit line usage or maturity.
[8] Following, for example, Ivashina and Scharfstein (2010b), if a credit line is syndicated, we control for the lead lender and identify it as the member of the syndicate designated as the administrative agent.

Regarding credit line characteristics, we control for the maturity stated in the credit contract and loan commitment and for whether the lines support the issuance of letters of credit, can be used for corporate restructuring, are provided by a syndicate of lenders, and are secured. We also control for whether facilities include commitment, annual, utilization, and upfront fees.

The reason to control for whether firms are in violation of the financial covenants of their credit lines is that these violations have been shown to have a significant impact on credit line-related features such as corporate investment (Chava and Roberts 2008, Nini et al. 2009), credit agreement renegotiations (Roberts and Sufi 2009, Roberts 2015), and credit line usage (Sufi 2009). Indeed, if financial covenants are breached and an actual event of technical default occurs, lenders could accelerate the credit line (Wight et al. 2007). We also control for the 2008 crisis, since previous works have observed that facility usage rose significantly during this period (Ivashina and Scharfstein 2010a). Moreover, we check whether the crisis modified the relation among facility usage, corporate creditworthiness, and pricing schedules.

### 4.2. Results

Columns (1) and (2) of Table 4 report the results from the estimation of the base regression model and from regressing the same model without the interaction terms between the Z-score and the indicator variables for the interest rate structures. As column (2) shows, the coefficient of the interaction term between the Z-score and the indicator variable for the xp lines ($\alpha_5 = -0.056$) is negative; that is, bearing in mind that the Z-score is an inverse measure of corporate financial distress, we find the effect of corporate financial distress on the utilization of a line of credit is statistically significantly more positive for firms with xp lines. Moreover, since the coefficient of the Z-score ($\alpha_4 = 0.033$) is smaller in absolute value than $\alpha_5$, xp lines are used more as the creditworthiness of the holding firms deteriorates. We do not observe, however, similar effects for the xx and pp lines, since the coefficients of the interaction terms between the Z-score and the indicator variables for the pp and xx lines ($\alpha_6$ and $\alpha_7$, respectively) are insignificant.

**[Table 4]**

Increases in a firm's financial distress are likely to lead to increases in its spot risk premium. Under these circumstances, firms would increasingly use debt instruments that protect them against rising financing costs. This is why, according to the hypothesis that fixed spreads provide interest rate insurance, we expect to observe debt instruments having a positive effect on how corporate financial distress relates with facility usage. In this sense, the fact that $\alpha_5$ is negative suggests that xp credit lines provide interest rate insurance to borrowers. Since performance-sensitive spreads are designed to reflect changes in risk premium, this protection is likely to be related to the option to borrow at the fixed spreads that xp lines offer and, hence, our results support the theoretical view of fixed spread credit lines as interest rate hedging instruments.

For similar reasons, we do not expect pp lines to have a positive effect on how usage changes as corporate creditworthiness deteriorates. Thus, the non-significance of the coefficient of the interaction term between the Z-score and the indicator variable for the pp lines ($\alpha_7$) is in line with our hypothesis that fixed spread credit facilities provide borrowers with interest rate insurance. Nevertheless, our results for xx lines are inconsistent with this hypothesis: As Table 4 shows, this type of credit line does not positively affect the relation between usage and financial distress.

Therefore, pp lines aside, there seem to be differences in how xp and xx lines are used as corporate creditworthiness changes. Since margins are fixed in xx lines but fixed and variable in xp lines, these differences highlight that the standard dichotomy between fixed and performance-based spreads is not enough to analyze credit line usage. Moreover, such differences underline the relevance of the fact that most credit lines offer borrowers the choice between ABR and LIBOR loans, although this feature has been ignored by previous research.

However, how can we explain that the usage of xx lines contradicts the hypothesis that credit lines with fixed spreads provide interest rate insurance? On the one hand, the probit analysis in the next section indicates that firms of uncertain or low quality are more likely to enter into xx lines. On the other hand, as Table 5 reports, the spreads of xx lines are surprisingly similar to the maximum spreads of pp lines, particularly for LIBOR loans. The means of the fixed and maximum spreads on the LIBOR loans of xx and pp lines are 263.11 basis points (bps) and 267.55 bps, respectively; the standard errors are 111.63 bps and 109.79 bps, respectively; and the medians are 250 bps for both types of credit lines. Indeed, both the Mann–Whitney test (Z = 0.009, p-value = 0.99) and the Kolmogorov–Smirnov test (D = 0.056, p-value = 0.97) do not reject the null hypothesis that the distribution of the fixed spreads on LIBOR loans in the xx lines and the distribution of the maximum spreads on LIBOR loans in the pp lines are drawn from the same population.

**[Table 5]**

These two findings suggest that lenders use xx lines to curb the incentives of uncertain or low-quality firms to borrow from credit lines. On average, xx lines not only are as costly as pp lines would be in the worst possible scenario, but also do not give firms the option to reduce these relatively high financing costs; that is, xx lines are an expensive debt instrument, regardless of how firms' creditworthiness evolves. In this sense, performance-based pricing aims at protecting lenders from risk shifting as corporate creditworthiness deteriorates. However, xx lines seem to pursue this same target, regardless of firms' financial condition. In this sense, if we test for the equality of the distribution functions of how all firms with xx lines use these facilities and how firms with pp lines that are in the lowest quintile of the Z-score use these credit instruments, the null hypothesis cannot be rejected. Specifically, the Mann–Whitney test (Z = 1.577, p-value = 0.115) and the Kolmogorov–Smirnov test (D = 0.077, p-value = 0.375) do not reject the null hypothesis that these two distributions are drawn from the same population.

The relation between credit line usage, interest rate structures, and corporate financial distress may be biased by endogeneity, in the sense that firms' choices among available credit lines could be based on private information about their future financial condition. However, some of our findings reduce concerns about the effects of this potential endogeneity bias over our results. First, as the analysis in the following section shows, future changes in firms' financial condition do not affect the probability of observing the different types of credit lines. Therefore, firms' private information on their level of future creditworthiness does not seem to determine the type of credit line into which firms enter. Second, endogeneity would lead us to expect a positive relation between firms' takedown probability and the probability of entering into credit lines that are used more intensively. Nevertheless, our results suggest the opposite. On the one hand, firms with low or uncertain quality and, hence, a higher takedown probability (Sufi 2009, Berg et al. 2016) are more likely to enter into xx lines. On the other hand, the results displayed on Table 4 indicate that that credit line usage decreases around 12% for lines with the xx interest rate structure. Therefore, despite endogeneity clearly being a relevant concern, we do not expect this potential problem to significantly affect our results.

The coefficient of the Z-score is significant only when the interaction terms between this variable and the dummies for the pricing schedules are included in the analysis. This result strengthens the idea that the analysis of whether revolving facilities are interest rate insurers can only be accurately carried out if interest rate structures are taken into account, because not all types of pricing schedule can provide such insurance.

Regarding firm characteristics, columns (1) and (2) of Table 4 show that, in tune with Berrospide et al. (2012), line usage is negatively related to firm size. This result is consistent with the idea that large firms have more financing sources available and, hence, can use lines of credit relatively less (Campello et al. 2011). The relation between liquidity and credit line usage is also negative. Highly liquid firms have a greater ability to turn their product into cash and, therefore, their liquidity management does not require a high rate of credit line usage. Indeed, the negative relation between liquidity and drawdowns is also observed by Acharya et al. (2014). Leverage has a positive effect on the utilization ratio. As for the relation between corporate size and line usage, such an effect can be explained as a result of the likely limited access of already highly leveraged firms to alternative financing sources and the use of drawdowns for debt repayment purposes. The limited and conditional access to financing sources also underlies the positive relation between the usage ratio and being a firm with no S&P corporate credit rating. Our results suggest that usage decreases with tangibility, but such an effect is only present when the interaction terms between the Z-score and the interest rate structures are included as regressors. With respect to credit line fees, usage appears to be higher only for facilities with annual fees.

In tune with Ivashina and Scharfstein (2010a), the 2018 crisis seems to have caused a significant increase in credit line mean usage: Our results suggest that the latter rose by around 10% during this period. In light of this result, we study whether the crisis changed the manner in which interest rate structures affect the relation between corporate creditworthiness and line usage. To perform this analysis, we include three additional interaction terms in the base model. Each of them examines how a given interest rate structure interacts with the Z-score and the indicator variable for the crisis. As column (3) of Table 4 shows, none of these interactions terms is significant; that is, the crisis seems to have no effect on how xp, xx, or pp lines are used as corporate quality deteriorates.

## 4.3. Probabilities of interest rate structures

This section studies the variables that affect the probability of observing interest rate structures. To fulfill this aim, we use maximum likelihood probit estimation. The relevant information for this analysis is constant or refers to the origination period and, therefore, the sample consists of one observation per facility. Given a firm's credit line, corporate characteristics refer to the quarter prior to the closing date of the line. Focusing on loan originations implies that the sample extends from 2001:Q2 to 2012:Q2. In this period, besides the 2008 crisis and according to the National Bureau of Economic Research (NBER), there is another crisis between 2001:Q1 and 2001:Q4. We modify the indicator variable for crisis to take this into account. We also add a linear time trend variable as a regressor. In addition, following Manso et al. (2010), we include a variable that allows us to examine whether the information that a firm has about its future financial distress affects the probability of observing the different types of credit lines. As a proxy for this information, for facility $i$ of firm $j$, we use a (0, 1) indicator variable that is equal to one if $j$'s Z-score is higher one year after $i$'s origination quarter than in the origination quarter and zero otherwise. Unless they predict perfect failure or success, regressions include one-digit SIC code industry and lead lender fixed effects.

**[Table 6]**

Table 6 presents the estimated coefficients and marginal effects from the three probit regressions performed, one for each of the interest rate structures. Column (2) indicates two strong predictors of the probability of a firm entering into an xx credit line: having no S&P corporate credit rating or being rated speculative grade. Holding all other variables at their median values, the probability of observing an xx line is 22.4% higher if the firm is not rated. The increase is even greater as firms go from being investment grade to speculative grade: The probability of observing xx lines increases by 77.6%. This result could help explain why the average spreads of xx lines are similar to the maximum spreads of pp lines: Lenders seem to fix spreads at relatively high, constant levels when they enter into a credit contract with firms of uncertain or low quality. No other firm characteristic seems to affect the probability of observing a given pricing schedule.

As Table 6 indicates, the worsening (or improvement) of the level of financial distress after facility origination does not affect the probability of observing any of the interest rate structures. This outcome suggests that information about potential future financial distress does not influence the type of credit line into which firms eventually enter.

Comparison of the marginal effects of the time trend variable in columns (1) and (3) of Table 6 indicates that the probability of observing xp lines decreases by 15.3% in 2006:Q2–2010:Q2, whereas that of observing pp lines increases by 18.3%. This result is consistent with the radical manner in which the market for lines of credit has changed. As Figure 1 shows, until the start of the 2008 crisis, the xp and pp lines have similar relative frequencies by quarter: On average, 34.44% and 39.34% of the credit lines present in each quarter between 2006:Q1 and 2007:Q3 inclusive are xp and pp lines, respectively. However, at the start of the crisis, the relative frequency of xp lines begins a continuous decline that accelerates after 2010:Q2. Therefore, the percentage of xp lines over the total number of sample lines is only 1.22% in 2012:Q2; that is, this type of interest rate structure seems to have virtually vanished from the market for revolving lines of credit. The evolution of the relative presence of pp lines in the sample inversely mirrors that of xp lines: The 30% decrease in the relative frequency of the latter becomes a 30% increase in the relative frequency of the former, which have a relative frequency of 67.07% in the last quarter of the sample period.

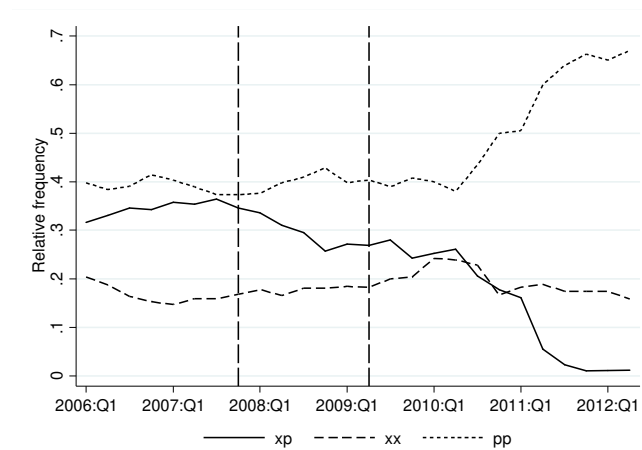

**Figure 1. Relative frequencies of the interest rate structures by quarter.** This figure represents the sample relative frequencies of the interest rate structures by quarter in 2006:Q1–2012:Q2. For an interest rate structure, the relative frequency by quarter is the ratio between the number of credit lines with this interest rate structure in a quarter and the total number of credit lines in that quarter. The vertical dashed lines mark the start (2007:Q4) and end (2009:Q2) of the 2008 crisis according to the NBER.

Regarding credit line characteristics, going from the lower to the higher limits of the interquartile ranges of credit line maturity and principal (i.e., from 27 months to 60 months and from $30 million to $350 million, respectively) reduces the probability of observing xx lines by 19.6% and 22.6%, respectively. Similarly, having a program to support the issuance of letters of credit and entering into a syndicated credit line reduces the probability of observing xx lines by 40.3% and 34.6%, respectively. In contrast to the results for xx lines, credit line maturity and principal and including a letter of credit program raises the probability of observing pp lines by 19%, 15%, and 17.6%, respectively. If loan contracts include utilization fees, the probability of observing pp lines decreases by 31.3%, whereas including commitment fees increases this probability by 24.4%. The probability of observing xp lines decreases by 11.5% if maturity increases and increases by 10.7% for syndicated lines of credit.

**4.4. Robustness checks**

To check for the robustness of our results in the base regression analysis, we perform a series of additional tests. First, instead of the Z-score, we use alternative proxies for corporate creditworthiness. Following Santos (2011), Roberts (2015), and Berg et al. (2016), we capture financial distress by equity volatility (EV), as measured by an ordered variable that groups firms into the quintiles of the 12-month standard deviation of their daily stock returns. The Center of Research in Security Prices (CRSP) US Stock Database provides required data to compute this standard deviation and the CRSP/Compustat Merged Database allows us to link such information to Compustat. In addition, we proxy for financial distress by means of Ohlson's (1980) O-score (OS) and a (0, 1) indicator variable that equals one if the firm is either rated speculative grade or has no S&P rating and its Z-score is below the sample median (SGNR). Second, we substitute the dependent variable in the base model by the non-averaged quarterly value of the usage-to-principal ratio. Third, we consider credit lines with spreads determined by more than one type of criterion as either xp or pp lines if one of those criteria is corporate performance. Fourth, we include the Kaplan–Zingales (1997) index (KZI) in the base analysis to control for firms that, despite having a credit line, are financially constrained in a more standard sense. In this regard, Sufi (2009) points out that these firms could use their credit lines differently. The definition of the KZI follows the linearization of Lamont et al. (2001). Fifth, given the potential biases associated with data on upfront fees, we check whether excluding the indicator variable for this fee has any effect on our results. In addition, since the credit line principal is part of the dependent variable and to exclude artificial correlations, we repeat the analysis without controlling for the size of the commitment.

[Table 7]

Taking into account that EV, OS, and SGNR are direct measures of corporate quality, Table 7 suggests that the results from our robustness tests are not qualitatively different from those of the base regression analysis.[9] In particular, except for the non-significance of the coefficient of SGNR, there is no qualitative change across specifications in the coefficient estimates of the indicator variables for the xp, xx, and pp lines; the variables that measure corporate financial distress; and the interaction terms between the former and latter variables. Regarding controlling for the level of corporate financial constraint, the coefficient estimate of the KZI is not statistically significant.

## 5. Conclusion

The theoretical view of fixed spread facilities as a hedging instrument against changes in spot corporate risk premiums has not been previously examined, probably due to lack of data on credit line usage at the facility level. This work is an attempt to fill this gap. In particular, we analyze the relation between credit line usage by mid- and small-cap–like firms and the interest rate structures of credit lines as corporate creditworthiness deteriorates. Typical credit lines offer borrowers the choice between ABR and LIBOR loans but the spreads added to each of these base rates are not necessarily of the same kind. The most common interest rate structures are those that offer borrowers fixed spreads on ABR loans and performance-sensitive spreads on LIBOR loans or the same kind of spreads, either fixed or performance sensitive, on both ABR and LIBOR loans. This plurality of interest rate structures suggests that the standard dichotomy of fixed versus variable spreads is insufficient for the analysis of the relation between usage and pricing.

In line with this complexity, not all fixed spread facilities seem to support the hypothesis that this type of spread provides interest rate insurance; specifically, the results regarding xx lines do not. Such a finding, however, is consistent with the spreads of these lines being fixed but at relatively high levels. Indeed, our results indicate that lenders seem to use xx pricing schedules to curb the incentives of uncertain or low-quality firms to pursue a risk-shifting strategy.

In contrast to what we observe for xx lines, xp lines appear to hedge against rising spot risk premiums and, therefore, these lines can be viewed as a risk-sharing mechanism used by borrowers to shift risk to lenders. Indeed, this risk shifting is a likely explanation for the dramatic decline in the presence of xp lines in the market since the 2008 crisis. As a result of this decline, the predominant forms of pricing in the market for revolving lines of credit have become pp and xx, that is, those where the spreads on both LIBOR and ABR loans are of the same kind, either flexible spreads that adapt to changes in firms' spot risk premiums or spreads fixed at relatively high levels.

---

[9] Table 7 only shows the estimated coefficients of the main variables of the analysis. Complete results are available upon request.

Regarding regulatory policy, as the Accords of Basel III and, previously, Basel II establish, determining the regulatory capital charge for credit risk requires an adequate estimation of exposure at default, that is, the loss that a financial intermediary is exposed to if a loan defaults. For lines of credit, computing such exposure implies estimating the percentage of the unused portion of the credit line that would be used in case of default. In this regard, previous works have pointed out that usage increases under default (Zhao et al. 2014). Our findings suggest that this increase may not be homogeneous across pricing schedules and, hence, the regulatory framework should take into account the potential effects of pricing schedules on firms' exposure at default.

**Appendix: Variable definitions**

**Interest rate structures**

- xp = 1 if the credit line allows fixed spread ABR loans and performance-sensitive spread LIBOR loans and 0 otherwise.

- xx = 1 if the credit line allows fixed spread ABR and LIBOR loans or just one of these types of loans is available and has fixed spreads and 0 otherwise.
- pp = 1 if the credit line allows performance-sensitive spread ABR and LIBOR loans or just one of these types of loans is available and has performance-sensitive spreads and 0 otherwise.

**Other loan characteristics**

- Usage-to-principal ratio = rolling four-quarter average of the available ratios of facility usage to loan commitment.
- Commitment fee = 1 if the credit line includes a fee charged on the unused amount of the commitment and 0 otherwise.
- Annual fee = 1 if the credit line includes a fee charged on the entire commitment amount, regardless of usage, and 0 otherwise.
- Utilization fee = 1 if the credit line includes a fee charged on the drawn amounts if and while a usage threshold is exceeded and 0 otherwise.
- Upfront fee = 1 if the credit line includes a single charge fee paid at origination and 0 otherwise.
- Letter of credit program = 1 if the credit line supports the issuance of letters of credit and 0 otherwise.
- Purpose = 1 if the credit line can be used for corporate restructuring, that is, leveraged buyouts, mergers and acquisitions, and stock repurchases, and 0 otherwise.
- Maturity = maturity specified in the credit contract.
- Syndicated line = 1 if the credit line is provided by a syndicate of lenders and 0 otherwise.
- Secured line = 1 if the credit line is secured and 0 otherwise.
- Principal = amount committed under the credit line.

**Firm characteristics**

The variables are defined in terms of their Compustat abbreviated names. If the name ends with $q$, it is a quarterly data item. Following Roberts and Sufi (2009) and except for rating-associated variables, all variables are equal to their rolling four-quarter averages, that is, the value of any variable in quarter $t$ is equal to its average between $t$ - 3 and $t$, inclusive. All financial variables from Compustat and the variable *ret*, taken from the CRSP database, are winsorized at the top and bottom percentiles. To facilitate the interpretation of coefficients, continuous variables used in interaction terms (Altman's Z-score and OS) are mean centered. This applies only to regressions where interaction terms are included as regressors; that is, it does not apply to descriptive statistics or probit regressions.

- Market-to-book ratio = (atq - (atq - ltq - pstkl + txditcq) + (cshoq * prccq))/atq.
- Leverage = (dlcq + dlttq)/atq.
- Liquidity = actq/lctq.
- Profitability = oibdpq/atq.
- Size = Ln(atq).
- Tangibility = ppentq/atq.
- Non-rated firm = 1 if the firm does not have an S&P corporate credit rating and 0 otherwise; S&P ratings are provided by the end-of-quarter value of splticrm.
- Speculative grade = 1 if the firm has an S&P corporate credit rating equal to BB+ or worse and 0 otherwise.
- Z-score = 1.2 * ((actq - lctq)/atq) + 1.4 * (req/atq) + 3.3 * (piq/atq) + 0.6 * ((prccq * cshoq)/ltq) + 0.999 * (saleq/atq).
- Change in the Z-score = Z-score$_{t+4}$ - Z-score$_t$, where $t$ is the origination quarter of the credit line.
- EV = ordered variable based on the quintiles of the 12-month standard deviation of ret, that is, of the firm's daily stock return.
- OS = - 1.32 - 0.407 * (log((atq/Deflator) * 100)) + 6.03 * (ltq/atq) - 1.43 * ((actq - lctq)/atq) + 0.076 * (lctq/actq) - 1.72 * X1 - 2.37 * (niq/atq) - 1.83 * ((piq + dpq)/ltq) + 0.285 * X2 - 0.521 * (niq – niq$_{t-1}$)/(|niq| + |niq$_{t-1}$|), where Deflator = quarterly, seasonally adjusted, implicit price deflator, as reported by the Federal Reserve of St. Louis; X1 = 1 if ltq ≥ atq and X1 = 0 otherwise; X2 = 1 if (niq + niq$_{t-1}$ + niq$_{t-2}$ + niq$_{t-3}$) < 0 or (niq$_{t-4}$ + niq$_{t-5}$ + niq$_{t-6}$ + niq$_{t-7}$) < 0 and X2 = 0 otherwise.
- SGNR = 1 if the firm is either rated speculative grade or has no S&P rating and its Z-score is below the sample median, and 0 otherwise.
- KZI = - 1.001909 * ((ibq + dpq)/ppentq$_{t-1}$) + 0.2826389 * (((atq - ceqq - txditcq) + (cshoq * prccq))/atq) + 3.139193 * ((dlcq + dlttq)/(dlcq + dlttq + seqq)) - 39.3678 * ((dvq + dvq$_{t-1}$ + dvq$_{t-2}$ + dvq$_{t-3}$)/ppentq$_{t-1}$) - 1.314759 * (cheq/ppentq$_{t-1}$).
- Covenant violation = 1 if the financial covenants of the facility are breached by the borrower and 0 otherwise.

## Table 1. Credit line characteristics

This table presents summary statistics for two samples. The first (usage sample) is the sample on which our analysis is based. The second sample (DealScan sample) consists of all the revolving credit lines in DealScan that are outstanding in the sample period, denominated in US dollars, and provided to non-financial US firms. To calculate the statistics referring to the usage sample, we proceed as if there were one observation per credit line. However, the statistics of the usage-to-principal ratio are calculated over the entire sample and the statistics of covenant violations are calculated as if there were one observation per firm. The variables are defined in the Appendix.

|  | Usage sample | | | DealScan sample | | |
|---|---|---|---|---|---|---|
|  | Mean | SE | Median | Mean | SE | Median |
| Usage-to-principal ratio | 0.24 | 0.28 | 0.10 |  |  |  |
| Letter of credit support (1, 0) | 0.86 | 0.34 | 1 |  |  |  |
| Purpose (1, 0) | 0.45 | 0.50 | 0 |  |  |  |
| Covenant violation (1, 0) | 0.21 | 0.41 | 0 |  |  |  |
| Maturity (months) | 41.87 | 19.05 | 46 | 50.08 | 18.72 | 60 |
| Principal ($mil) | 251.05 | 330.00 | 107.5 | 266.09 | 596.79 | 90 |
| Syndicated line (1, 0) | 0.92 | 0.28 | 1 | 0.99 | 0.12 | 1 |
| Secured line (1, 0) | 0.63 | 0.48 | 1 | 0.79 | 0.41 | 1 |
|  |  |  |  |  |  |  |
| Credit lines | 586 | | | 23,148 | | |
| Firms | 139 | | | 10,191 | | |

## Table 2. Firm characteristics

This table presents summary statistics for two samples. The first (usage sample) is the sample on which our analysis is based. The second sample (Compustat sample) consists of all firm–quarter observations from non-financial US firms appearing in Compustat between 2006:Q1 and 2012:Q2. For comparability with our sample, the firm characteristics in the Compustat sample are calculated as rolling four-quarter averages. The variables are defined in the Appendix.

|  | Usage sample | | | Compustat sample | | |
|---|---|---|---|---|---|---|
|  | Mean | SE | Median | Mean | SE | Median |
| Assets ($Mil) | 2,230.51 | 3,378.26 | 782.25 | 2,832.65 | 10,556.09 | 237.86 |
| Market-to-book ratio | 1.66 | 1.02 | 1.35 | 16.96 | 251.07 | 1.63 |
| Tangibility | 0.29 | 0.24 | 0.21 | 0.25 | 0.25 | 0.15 |
| Profitability | 0.03 | 0.03 | 0.03 | -0.20 | 3.68 | 0.02 |
| Leverage | 0.29 | 0.28 | 0.25 | 0.92 | 11.71 | 0.20 |
| Liquidity | 1.90 | 1.13 | 1.65 | 3.49 | 16.27 | 1.85 |
| Non-rated firm (1, 0) | 0.58 | 0.49 | 1 | 0.76 | 0.43 | 1 |
| Speculative grade (1, 0) | 0.27 | 0.44 | 0 | 0.14 | 0.34 | 0 |
| Altman's Z-score | 2.04 | 2.50 | 1.75 | -30.25 | 447.73 | 1.70 |
|  |  |  |  |  |  |  |
| Line–quarter obs. | 2,595 | | | 131,579 | | |
| Firms | 139 | | | 7,031 | | |

## Table 3. Interest rate structures

This table presents the types of interest rate structures and fees in our sample. Panel A shows the types of spreads of sample credit lines for ABR and LIBOR base rates. Panel B reports the distribution of sample lines across interest rate structures. To name the interest rate structures, we use two letters: the first (second) one refers to the type of spread, if any, over ABR (LIBOR) loans. The letters used are *x* (fixed spreads), *p* (performance-sensitive spreads), *u* (usage-based spreads), and *m* (maturity-based spreads). The columns in Panel B referring to multiple criteria provide information about lines whose spreads are determined by more than a single criterion. Here, *usg* stands for usage-based spreads, *prf* for performance-sensitive spreads, and *mat* for maturity-based spreads. Accordingly, the column *usg & prf*, for instance, refers to lines whose spreads are jointly determined by firm performance and line usage and the subdivision between *LIBOR* and *Both* refers to whether the LIBOR base rate or both the ABR and LIBOR base rates are, respectively, determined by more than one criterion. The percentages in Panel A are computed over the total number of credit lines appearing in the bottom cell of each column. The percentages in Panels B are relative to the total number of sample credit lines.

| Panel A. Base rates and fees | | |
|---|---|---|
| Types of spreads | Types of base rates | |
| | ABR | LIBOR |
| Performance sensitive | 280 (49.82%) | 392 (70.38%) |
| Fixed | 226 (40.21%) | 104 (18.67%) |
| Usage based | 46 (8.19%) | 48 (8.62%) |
| Maturity based | 3 (0.53%) | 1 (0.18%) |
| Multiple criteria | 7 (1.25%) | 12 (2.15%) |
| Total | 562 | 557 |

| Panel B. Types of interest rate structures | | | | | | | | |
|---|---|---|---|---|---|---|---|---|
| pp | xx | xp | uu | mm | xu | Multiple criteria | | |
| | | | | | | | LIBOR | Both |
| | | | | | | usg & prf | 5 (0.85%) | 2 (0.34%) |
| 286 (48.81%) | 120 (20.48%) | 115 (19.62%) | 46 (7.85%) | 3 (0.51%) | 2 (0.34%) | usg & mat | 0 | 4 (0.68%) |
| | | | | | | usg & prf & mat | 0 | 1 (0.17%) |

**Table 4. Interest rate structures, usage, and financial distress**

This table presents the results from regressions that analyze the effect of the interest rate structure on the relation between facility usage and corporate financial distress. All variables referring to firm characteristics are lagged by one quarter. Column (1) reports the coefficient estimates of the base regression model without interaction terms. Column (2) reports the coefficient estimates of the base regression model. Column (3) analyzes the effect of the 2008 crisis on the relation among usage, pricing schedules, and financial distress. Regressions include quarter, one-digit SIC industry, and lead lender fixed effects. Statistical significance at the 5% and 1% levels is denoted by * and **, respectively. All regressions are conducted with standard errors robust to within-line dependence and heteroskedasticity. The variables are defined in the Appendix.

| | Interest rate structures (1) | Financial distress (2) | Crisis (3) |
|---|---|---|---|
| Intercept | 0.620** (0.189) | 0.623** (0.182) | 0.623** (0.184) |
| xp | 0.002 (0.050) | 0.068 (0.055) | 0.068 (0.055) |
| xx | -0.126** (0.046) | -0.118* (0.046) | -0.118* (0.046) |
| pp | -0.046 (0.038) | -0.022 (0.040) | -0.022 (0.040) |
| Z-score | 0.004 (0.008) | 0.031* (0.014) | 0.031* (0.014) |
| Z-score*xp | - | -0.059** (0.018) | -0.060** (0.018) |
| Z-score*xx | - | -0.023 (0.014) | -0.022 (0.014) |
| Z-score*pp | - | -0.008 (0.016) | -0.008 (0.016) |
| Crisis | 0.107** | 0.100** | 0.098** |

| | (0.035) | (0.034) | (0.035) |
|---|---|---|---|
| Z-score*Crisis*xp | - | - | 0.003 (0.014) |
| Z-score*Crisis*xx | - | - | -0.004 (0.008) |
| Z-score*Crisis*pp | - | - | -0.001 (0.014) |
| Size | -0.076** (0.017) | -0.072** (0.018) | -0.072** (0.018) |
| Market-to-book ratio | -0.030 (0.016) | -0.024 (0.017) | -0.024 (0.017) |
| Tangibility | -0.125 (0.067) | -0.137* (0.067) | -0.138* (0.068) |
| Profitability | -0.506 (0.423) | -0.843 (0.449) | -0.846 (0.448) |
| Leverage | 0.529** (0.090) | 0.544** (0.091) | 0.545** (0.091) |
| Liquidity | -0.034** (0.011) | -0.041** (0.012) | -0.041** (0.012) |
| Non-rated firm | 0.183* (0.071) | 0.193** (0.069) | 0.194** (0.069) |
| Speculative grade | 0.076 (0.052) | 0.095 (0.052) | 0.096 (0.052) |
| Commitment fee | 0.126 (0.073) | 0.135 (0.071) | 0.135 (0.071) |
| Annual fee | 0.167* (0.078) | 0.178* (0.074) | 0.178* (0.074) |
| Utilization fee | 0.027 (0.064) | 0.040 (0.063) | 0.040 (0.063) |
| Upfront fee | -0.009 (0.027) | 0.000 (0.026) | 0.000 (0.027) |
| Letter of credit program | -0.025 (0.043) | -0.014 (0.041) | -0.014 (0.041) |
| Purpose | -0.032 (0.027) | -0.030 (0.026) | -0.031 (0.026) |
| Maturity | -0.000 (0.001) | -0.000 (0.001) | -0.000 (0.001) |
| Principal | 0.000 (0.000) | 0.000 (0.000) | 0.000 (0.000) |
| Syndicated line | 0.009 (0.044) | -0.016 (0.043) | -0.017 (0.042) |
| Secured line | -0.043 (0.031) | -0.035 (0.030) | -0.035 (0.030) |
| Covenant violation | -0.010 (0.043) | -0.011 (0.040) | -0.014 (0.041) |
| Lines of credit | 485 | 485 | 485 |
| Observations | 1,956 | 1,956 | 1,956 |
| $R^2$ | 0.494 | 0.512 | 0.512 |

### Table 5. Maximum, minimum, or fixed values of spreads

This table presents summary statistics of the value (for fixed spreads) or the maximum and minimum (for performance-sensitive spreads) of the spreads on LIBOR and ABR loans across the main interest rate structures of the sample lines. The interest rate structures are defined in the Appendix.

| Spreads | | xp | | | xx | | | pp | | |
|---|---|---|---|---|---|---|---|---|---|---|
| | | Mean | SE | Median | Mean | SE | Median | Mean | SE | Median |
| LIBOR loans | Max. or fixed | 157.24 | 88.03 | 150 | 263.11 | 111.63 | 250 | 267.55 | 109.79 | 250 |
| | Min. or fixed | 99.67 | 79.06 | 75 | | | | 171.11 | 97.64 | 150 |

| ABR loans | Max. or fixed | 10.00 | 72.38 | 0 | 102.52 | 133.77 | 50 | 145.75 | 114.78 | 125 |
|---|---|---|---|---|---|---|---|---|---|---|
|  | Min. or fixed |  |  |  |  |  |  | 63.15 | 97.79 | 25 |

**Table 6. Determinants of the probabilities of the interest rate structures**

This table presents the estimated coefficients, marginal effects, and standard errors (in parentheses) from three probit regressions. The dependent variable of each of the regressions is a (0, 1) indicator variable for one of the three main interest rate structures. The marginal effect of each variable is computed as the effect over the predicted probabilities of going from the 25th to the 75th percentile of the variable distribution for a continuous variable and of going from zero to one for an indicator variable, holding all other variables constant at their median values. The sample for this analysis includes only one observation per line of credit. Firm characteristics correspond to the quarter prior to credit line origination. The indicator variable for crisis is one if the origination took place in 2001:Q2–2001:Q4 or 2007:Q4–2009:Q2 and zero otherwise. A linear time trend variable and a (0, 1) indicator variable that captures whether firms' financial distress improves or worsens after origination are included as regressors. The (0, 1) indicator variables for annual and utilization fees are not included in the regression corresponding to xx lines because these variables perfectly predict failure in the dependent variable. Regressions include one-digit SIC industry and lead lender (0, 1) indicator variables, which, however, are not used if they perfectly predict failure or success. Statistical significance at the 5% and 1% levels is denoted by * and **, respectively. All regressions are conducted with standard errors robust to within-firm dependence and heteroskedasticity. The variables are defined in the Appendix.

|  | xp (1) | | xx (2) | | pp (3) | |
|---|---|---|---|---|---|---|
|  | Coefficient | Marginal effect | Coefficient | Marginal effect | Coefficient | Marginal effect |
| Z-score | 0.163 (0.116) | 0.035 | 0.016 (0.101) | 0.009 | 0.135 (0.078) | 0.092 |
| $\Delta$(Z-score) | -0.649 (0.400) | -0.050 | 0.118 (0.259) | 0.030 | 0.293 (0.224) | 0.091 |
| Crisis | -0.267 (0.345) | -0.023 | -0.112 (0.250) | -0.026 | 0.270 (0.224) | 0.084 |
| Ln(Assets) | -0.309 (0.197) | -0.074 | 0.194 (0.170) | 0.117 | -0.009 (0.141) | -0.007 |
| Market-to-book ratio | -0.317 (0.224) | -0.024 | -0.203 (0.128) | -0.040 | -0.248 (0.137) | -0.063 |
| Tangibility | 0.428 (0.815) | 0.014 | -0.997 (0.710) | -0.078 | 0.711 (0.764) | 0.073 |
| Profitability | -4.357 (4.525) | -0.009 | -1.290 (3.026) | -0.006 | 3.706 (2.962) | 0.024 |
| Leverage | -1.476 (1.155) | -0.039 | 1.013 (0.916) | 0.068 | 0.380 (0.668) | 0.033 |
| Liquidity | 0.374 (0.204) | 0.048 | -0.119 (0.178) | -0.037 | -0.228 (0.153) | -0.091 |
| Non-rated firm | 0.325 (0.745) | 0.028 | 7.349** (1.224) | 0.224** | -0.317 (0.563) | -0.098 |
| Speculative grade | 0.392 (0.676) | 0.042 | 7.279** (1.148) | 0.776** | -0.399 (0.531) | -0.120 |
| Commitment fee | -1.447 (0.767) | -0.212 | 0.218 (0.458) | 0.049 | 0.885* (0.394) | 0.244* |
| Annual fee | 1.321 (0.841) | 0.188 | - | - | 0.549 (0.482) | 0.168 |
| Utilization fee | 0.548 (0.591) | 0.062 | - | - | -1.230* (0.597) | -0.313* |
| Letter of credit program | 0.453 (0.462) | 0.037 | -1.298** (0.368) | -0.403** | 0.604* (0.308) | 0.176* |
| Purpose | -0.139 (0.313) | -0.013 | -0.376 (0.295) | -0.080 | 0.171 (0.243) | 0.053 |
| Maturity | -0.033* (0.013) | -0.115* | -0.023** (0.007) | -0.196** | 0.019** (0.006) | 0.190** |
| Principal | 0.000 (0.001) | 0.007 | -0.004** (0.001) | -0.226** | 0.002* (0.001) | 0.150* |
| Syndicated line | 2.647** (0.574) | 0.107** | -1.124* (0.452) | -0.346* | 0.892 (0.470) | 0.246 |
| Secured line | -0.263 (0.385) | -0.027 | 0.350 (0.321) | 0.075 | 0.074 (0.292) | 0.023 |
| Time trend | -0.104** (0.022) | -0.153** | 0.011 (0.019) | 0.044 | 0.037** (0.013) | 0.183** |
| Credit lines | 355 | | 355 | | 355 | |
| Prob(Y=1) | 0.111 | | 0.224 | | 0.464 | |
| Pseudo-$R^2$ | 0.587 | | 0.452 | | 0.309 | |

### Table 7. Robustness checks

This table presents the results from robustness checks of the analysis of the relation among interest rate structures, corporate financial distress, and facility usage. Columns (1) and (2) show the coefficient estimates from analyses where the variable used to measure financial distress (Finc. dist.) is not Altman's Z-score. This variable is substituted by EV (i.e., equity volatility) in column (1), OS (i.e., Ohlson's O-score) in column (2) and SGNR (i.e., an indicator variable that equals one if the firm is either rated speculative grade or has no S&P rating and its Z-score is below the sample median) in column (3). In column (4), the dependent variable (i.e., the usage-to-principal ratio) is not averaged over the last four quarters. In column (5), xp and pp stand for credit lines whose variable spreads are jointly determined by corporate performance and other criteria. Column (6) includes the KZI (i.e., the Kaplan–Zingales index) as a control variable. Columns (7) and (8) do not include (0, 1) indicator variables for upfront fees and credit line principal, respectively. Regressions include quarter, one-digit SIC industry, and lead lender fixed effects. Statistical significance at the 5% and 1% levels is denoted by * and **, respectively. All regressions are conducted with standard errors robust to within-line dependence and heteroskedasticity. The variables are defined in Appendix A.

| | EV (1) | OS (2) | SN (3) | Not avg. (4) | Mult. crit. (5) | KZI (6) | Upf. fee (7) | Principal (8) |
|---|---|---|---|---|---|---|---|---|
| xp | -0.117 (0.087) | 0.031 (0.059) | -0.066 (0.055) | 0.075 (0.055) | 0.056 (0.058) | 0.070 (0.055) | 0.068 (0.055) | 0.073 (0.054) |
| xx | -0.227* (0.090) | -0.136** (0.046) | -0.167** (0.056) | -0.111* (0.047) | -0.128** (0.047) | -0.113* (0.047) | -0.118* (0.045) | -0.117** (0.045) |
| pp | -0.125 (0.067) | -0.041 (0.038) | -0.038 (0.048) | -0.016 (0.040) | -0.034 (0.044) | -0.020 (0.040) | -0.022 (0.039) | -0.017 (0.039) |
| Finc. dist. | -0.055** (0.018) | -0.070** (0.015) | -0.036 (0.060) | 0.033* (0.014) | 0.034* (0.016) | 0.029* (0.014) | 0.031* (0.014) | 0.032* (0.014) |
| Finc. dist.*xp | 0.049* (0.023) | 0.052* (0.024) | 0.200* (0.081) | -0.056** (0.019) | -0.063** (0.020) | -0.057** (0.018) | -0.059** (0.018) | -0.062** (0.018) |
| Finc. dist.*xx | 0.040 (0.023) | 0.029 (0.019) | 0.089 (0.067) | -0.023 (0.014) | -0.027 (0.016) | -0.024 (0.014) | -0.023 (0.014) | -0.025 (0.014) |
| Finc. dist.*pp | 0.031 (0.019) | 0.011 (0.018) | 0.023 (0.063) | -0.005 (0.016) | -0.012 (0.018) | -0.006 (0.016) | -0.008 (0.016) | -0.010 (0.016) |
| KZI | - | - | - | - | - | 0.000 (0.000) | - | - |
| Obs. | 1,980 | 1,949 | 2,045 | 1,956 | 1,926 | 1,916 | 1,956 | 1,956 |
| $R^2$ | 0.514 | 0.506 | 0.487 | 0.462 | 0.509 | 0.507 | 0.512 | 0.511 |

**Highlights**

- This paper tests whether fixed spread credit lines provide interest rate insurance.
- The research is performed at the facility level with hand-collected data.
- Interest rate structures are at the core of the analysis.
- Not all structures with fixed spreads hedge against rising risk premiums.
- The dichotomy fixed-variable spread is inaccurate to examine facility pricing.

## Response to reviewers
## "Pricing and Usage: An Empirical Analysis of Lines of Credit"

Thank you very much for your revision suggestions: They have helped us to enhance the quality of the paper. Your comments appear in bold below, followed by point-by-point explanations of how we have tried to address your concerns. We have also included indications, when necessary, of where the revisions can be found in the new version of the paper.

### Reviewer #1

**This is an interesting and well-executed paper that is likely to be of significant interest to the journal's readers. I am happy to recommend it for publication subject to the following:**

**1. My first issue is the selection of Altman Z-score and Ohlson O-score. While both are fixed linear models that aim to proxy the creditworthiness of borrowers, certain lenders may in fact use more modern/advanced models such as term-structure derivation of credit risk or RAROC to determine risk premiums and potentially the type of interest rate structure offered to the lender.**

As Reviewer#1 points out, the term structure derivation of credit risk (TSDCR) and the risk-adjusted return on capital (RAROC) are sophisticated methods for evaluating and pricing credit risk. They provide lenders with a risk-based method to assess economic performance, that is, whether the economic return on assets is commensurate with their risk. Nevertheless, the focus of our paper is not on lenders but borrowers, in particular, on how borrowers' financial distress affects facility usage given an interest rate structure. Although the lender and borrower perspectives are closely related, the fact that we focus on the latter implies that the TSDCR and the RAROC may not be, in our view, adequate for capturing corporate financial distress in our analysis.

In this regard, we would like to point out the following. First, the RAROC would be close to zero (an indication of low performance for the lender, i.e., high risk) if a line of credit is not used, that is, if there are no borrowings outstanding. The reason is that, although some fees (such as annual or commitment fees) would be charged, interest rates on borrowings, which are facilities' main source of income, would yield no return if there are no drawdowns.[10] Nevertheless, despite having a RAROC close to zero, the firm that holds the line of credit can be high quality. Indeed, high-quality firms tend to use their facilities less (Sufi 2009, Berg et al. 2016).

---
[10] Even for a risk-free loan, the denominator of RAROC is not zero if the facility is not used (and the maturity is longer than one year): Lenders must set aside regulatory capital to cover contingent exposure to the unused portion of the commitment. In this regard, see Section D, Conversion Factors for Off-Balance

Second, net income on lines of credit (i.e., the numerator of the RAROC) depends on usage, which is our dependent variable. This feature of our model suggests that using the RAROC as the variable that captures corporate creditworthiness is not desirable.

Third, the data required to compute the RAROC are not available: Except for the origination quarter, there is no public information on the concrete interest rates charged on facilities with performance-sensitive spreads.

Fourth, the basic idea behind the TSDCR is that the risk premium of a loan can be derived from the distance between the yield curves of the risk-free asset and zero-coupon bonds with the same rating as the loan. Nevertheless, on the one hand, more than half of the sample firms (58%) are not rated. On the other hand, while defining risk premiums by the category of firms is adequate for lenders' risk management, such a grouping can imply a loss of accuracy for our analysis.

Despite their limitations, the Z- and O-scores are widely accepted proxies of corporate financial distress. The former, for instance, is the measure used by Sufi (2009), one of the most outstanding papers in the empirical analysis of the market for lines of credit.

Nevertheless, given Reviewer #1's concerns on the use of the Z- and O-scores to measure corporate financial distress, we perform an additional robustness test in which we use an alternative variable. In this test, following Santos (2011), Roberts (2015), and Berg et al. (2016), we capture financial distress by equity volatility. Specifically, we use an ordered variable based on the quintiles of the 12-month standard deviation of firms' daily stock returns. The results of this robustness check are qualitatively similar to those obtained from the base regression analysis. In particular, there is no qualitative change in the coefficient estimates of the indicator variables for the interest rate structures, the variables that measure corporate financial distress, and the interaction terms between the former and latter variables (*see column (1) in Table 7 and Section 4.4*).

Table 7 in the paper displays only the results for the main variables of our analysis. Therefore, we attach a document that shows all the estimated coefficients when financial distress is measured by means of equity volatility.

---

Sheet Items, of the Federal Deposit Insurance Corporation Rules and Regulations, Appendix A to Part 325, Statement of Policy on Risk-Based Capital.

**2. Furthermore, the risk premium and type of interest rate structure may be depending on certain characteristics of the already existing loan portfolio of the lender.**

**3. The methodology is well developed. However, a question that arises is whether the type of interest rate structure offered is: a) dependent on the lender and b) dependent on the creditworthiness/characteristics of the borrower. As such, the study may benefit from a lender specific fixed-effect.**

In our view, points 2 and 3 can be addressed together. To control for potential effects that lenders can have on the relations among credit line usage, interest rate structures, and corporate financial distress, we have included lender fixed effects in our empirical analysis (in the base model, robustness checks, and probit analysis). Following Ivashina and Scharfstein (2010b), if a credit line is syndicated we control for the lead lender and identify it as the member of the syndicate designated as the administrative agent (*see footnote 9, Section 4.1*).

Regarding our main results, taking into account lead lender fixed effects does not have qualitative effects. However, it affects control variables in an interesting way. For instance, in tune with previous findings (Ivashina and Scharfstein 2010a) and in contrast to the former version of the paper, our results indicate that facility usage increased during the 2008 crisis (*see Table 4 and the last paragraph of Section 4.2*).

**4. The analysis presented in the paper is interesting and sound. However, the body of the paper would benefit from further discussion of the results and potentially international institutional context.**

We have tried to improve the discussion of the results throughout the paper (*see Sections 1, 4.2, 4.3, and 5*).

To provide the international institutional context of our analysis, we have highlighted the international nature of the market for lines of credit (*see the first paragraph of Section 1*).

**5. Please ensure the reference list is fully up to date and formatting is consistent (e.g. check italic letters).**

We have checked that every reference cited in the text is also present in the reference list and vice versa. We have also revised the reference formatting for consistency; indeed, we have adapted it to the JIFMIM's style guidelines.

**6. Please proof read carefully, there remain typos, grammatical errors in the text.**

The paper has been re-edited by a company specialized in academic editing and translation.

**7. Please ensure the paper fully meets JIFMIM's style guidelines.**

We have modified the paper so that it satisfies these guidelines.

**Reviewer #2**

**The present study examines the relationship between credit line usage and corporate creditworthiness given the interest rate structure of credit lines (i.e. fixed and performance sensitive spreads or rates linked to the corporate performance). The paper uses data at credit line level from mid and small-cap US publicly traded corporations that have at least one active credit line in the sample period q1 2006 - q2 2012, thus including the financial crisis effects.**

**The derived results show that firms with credit lines with fixed and performance sensitive spreads use them more as they become increasingly financially distressed. Moreover, the paper investigates the variables that affect the probability of observing the different interest rate structures through the use of a maximum likelihood probit estimation.**

**The paper is well structured. The title describes clearly the paper while the abstract reflects fully the content of the article. Moreover, the results are clearly laid out with logical sequence.**

**Generally, I like the idea of using credit line level data to test the relationship between interest rate structures and usage. In this direction, the authors seem to consumed a lot of time for gathering and processing it, showing at the same time and professional experience given the identification of specific credit line characteristics such as the use of credit lines for the issuance of letters of credit. However, I have some concerns with this study that I outline below:**

**1. The main major concern is related with the sampling process. Although the final data set (after adjustments) has 2.545 firms, the authors end up (with a random process) with only 139 firms (and approximate the same number of credit lines). The authors should expand more the sample data set (given the population of 2.545 firms) in order to increase the validity of their results.**

We are fully aware that our sample size is limited. The main reason for this is that most of the data on lines of credit has been hand-collected in a process that has involved a large amount of time and substantial effort.

In this respect, we have manually gathered information for more than 586 credit contracts (the total number of facilities in our final sample) on aspects such as fees, principal, maturity, method of syndication, secured/unsecured, purpose, issuance of letters of credit, base rates, and spreads. In addition, we have obtained data on facility usage and covenant violations for a 26-quarter period. Moreover, we have checked a large number (around 150 randomly selected firms times 26 quarters) of 10-K and 10-Q SEC reports to find credit contracts not covered by DealScan (for more details, *see Section 3.2*).

This search process lasted about 12 months. The result is a sample with 2,595 observations. Despite its limitations, the following should be taken into account: First, the unit of analysis of our investigation is not the firm but the line of credit (*see the base model in Section 4.1*). In this regard, our sample includes 586 facilities. Second, key papers in this area of research that have also hand-collected information use datasets with a number of observations close to ours. In this sense, for instance, the hand-collected samples of Sufi (2009) and Roberts (2015) include 1,908 and 1,083 observations, respectively.

**2. The theoretical motivation and hypothesis Section should include more literature related to the theory and empirical findings of interest rate structures which is the main research topic of the study. Also, I propose the theoretical part as regards the pricing of revolving credit lines (pages 5-7) to be limited. Since the authors devote only one paragraph (at page 8) of this section for developing a theory relevant hypothesis, I suggest this part to be redrafted and be focused on one or two hypotheses with the credible nulls that can be tested with the expanded unique data set.**

We have completely rewritten Section 2. First, we have included more empirical literature on pricing schedules in lines of credit. Second, we have limited the discussion of the theoretical motivation. Third, closely related to the previous point and following the JIFMIM's guidelines, we have avoided extensive discussion of previous works. Fourth, we have explicitly included the hypothesis of our analysis, along with a brief discussion of its foundations.

**3. As regards the empirical model and given that the occurrence of financial crisis affected the credit line available balances (i.e. in some cases banks cut off the available balance of credit lines), the authors should focus more on the crisis effects. Thus, it would be interesting to examine the relationship between credit line usage**

**and a given interest rate structure when the borrower's creditworthiness changes and the environment changes too (i.e. interaction term between interest rate structure, crisis variable and borrower's creditworthiness).**

Following this suggestion, we included three additional interaction terms in the base model. Each of them examines how a given interest rate structure (xp, xx, or pp) interacts with financial distress and an indicator variable for the 2008 crisis. Nevertheless, none of these interactions terms is statistically significant; that is, the crisis seems to have no effect on how xp, xx, or pp lines are used as corporate quality deteriorates (*see column (3) of Table 4 and the last paragraph of Section 4.2*).

Despite this result, the crisis appears to have led to a substantial increase in facility usage (*see the last paragraph of our answers to points 1 and 2 in response to Referee #1*).

**4. The conclusion section need to be tightened. I suggest to split the section in two parts, one that summarizes the main findings in line with your hypotheses (see comment 2 above) and one that offers policy implications.**

We have rewritten the conclusion (*see Section 5*). First, we have implicitly split Section 5 into two parts. Second, the first part puts forward, in a summarized manner, the key features of our research and our main findings. Third, these findings are discussed in relation to the hypothesis in Section 2. Fourth, we discuss the policy implications of our results.

Regarding the last point, one of our main conclusions is that the relation between facility usage and financial distress is not homogeneous across pricing schedules. This result suggests that exposure at default can depend on the interest rate structure of lines and credit. Hence, given the relevance that exposure at default has in Basel II and III in relation to determining the capital charge on credit risk, the regulatory framework should take into account potential effects of pricing on credit risk (*see the last paragraph of Section 5*).

We hope to have completed all your requirements in a satisfactory manner. However, we are willing to accommodate any further concerns you may have.

Once again, thank you very much for helping us to improve the paper.